\documentclass[prl,aps,showpacs,twocolumn,amssymb,amsmath,amstext,amsfont,noshowkeys]{revtex4}
\usepackage{graphicx}
\usepackage{theorem}
\usepackage{dcolumn}
\usepackage{bm}

\begin{document}

\title{Metastable helium molecules as tracers in superfluid liquid $^{4}$He}
\author{W. Guo}\author{J.D. Wright}\author{S.B. Cahn}\author{J.A. Nikkel} \author{D.N. McKinsey}
\affiliation{Physics department, Yale University, New Haven, CT 06515}
\date{\today}

\begin{abstract}
Metastable helium molecules generated in a discharge near a sharp tungsten tip operated in either
pulsed mode or continuous field-emission mode in superfluid liquid $^{4}$He are imaged using a
laser-induced-fluorescence technique. By pulsing the tip, a small cloud of He$_{2}^{*}$ molecules
is produced. At 2.0~K, the molecules in the liquid follow the motion of the normal fluid. We can
determine the normal-fluid velocity in a heat-induced counterflow by tracing the position of a
single molecule cloud. As we run the tip in continuous field-emission mode, a normal-fluid jet from
the tip is generated and molecules are entrained in the jet. A focused 910~nm pump laser pulse is
used to drive a small group of molecules to the vibrational $a(1)$ state. Subsequent imaging of the
tagged $a(1)$ molecules with an expanded 925~nm probe laser pulse allows us to measure the velocity
of the normal fluid. The techniques we developed demonstrate for the first time the ability to
trace the normal-fluid component in superfluid helium using angstrom-sized particles.
\end{abstract}

 \maketitle

Visualizing the flow of superfluid $^{4}$He has long been of interest to the scientific community
\cite{Chung,MurakamiCryo}. Recently, particle image velocimetry with polymer micro-spheres and
hydrogen isotopes has been used to study liquid helium flows \cite{Sciver JLTP 2005,Sciver Nat
phys} and solid hydrogen tracers have been used to visualize the quantized vortices
\cite{Bewley,Paoletti PRL}. However, the dynamics of micron-sized tracers in the presence of
vortices are complex~\cite{Sciver JLTP 2005}. Kivotides~\cite{Kivotides2008-78} analyzed the
results of Zhang and Van Sciver~\cite{Sciver JLTP 2005} and concluded that one must account for
particle-vortex interactions~\cite{Kivotides2008} in order to extract an accurate measurement of
the local normal-fluid velocity from experimental data obtained using micron-sized tracer
particles. Furthermore, if the vortex-line density is too high then the possibility to use
micron-sized particles to measure the normal-fluid velocity is lost. On the other hand, it has been
shown that metastable He$_{2}^{*}$ triplet molecules can be imaged using a
laser-induced-fluorescence technique \cite{McKinsey,Rellergert}. The He$_{2}^{*}$ molecules are
much smaller in size (7~{\AA} radius \cite{Benderskii}) and should follow the motion of the normal
fluid without being affected by vortices at temperatures above 1~K \cite{Vinen}. Although so far
the sensitivity in imaging the molecules is not high enough to track the motions of individual
molecules, useful studies can still be performed by tracking a group of molecules.~\cite{Rellergert
thesis} In this Letter we shall show in two demonstration experiments the methods we developed in
tracing the true normal-fluid flow. In the first demonstration experiment, a cloud of He$_{2}^{*}$
molecules was used as a single tracer. In the second experiment, a small group of He$_{2}^{*}$
molecules was tagged and imaged using their internal vibrational levels.

\begin{figure}
\includegraphics[scale=0.33]{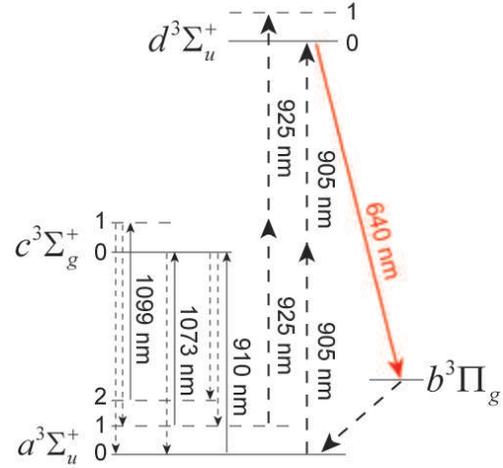}
\caption{Schematic diagram showing the cycling transitions for imaging the He$_{2}^{*}$ triplet
molecules.}\label{laser}
\end{figure}
Both experiments were conducted at 2.0~K. A sharp tungsten tip, made via a standard chemical
etching technique~\cite{Golov}, was used to produce the He$_{2}^{*}$ molecules in liquid helium. It
is known that He$_{2}^{*}$ molecules in both spin singlet and triplet states are produced near the
tip apex when a negative voltage with amplitude higher than the field-emission threshold is applied
to the tip \cite{McClintock,Dahm}. The singlet molecules radiatively decay in a few nanoseconds
\cite{Chabalowski}, while the triplet molecules are metastable with a radiative lifetime of about
13~s in liquid $^{4}$He \cite{McKinsey1999}. The widths of the He$_{2}^{*}$ molecule absorption
spectral lines in liquid helium (120~cm$^{-1}$~\cite{Hill}) are considerably larger than the
spacings of the rotational levels ($\sim$7~cm$^{-1}$~\cite{Herzberg}). A single pulsed laser at
905~nm is able to drive triplet molecules out of the $a^{3}\Sigma_{u}^{+}$ state to produce
fluorescence through a cycling transition (see Fig.~\ref{laser}) \cite{Rellergert}. However, the
vibrational levels are separated by about 1500~cm$^{-1}$ \cite{Herzberg}, and the
vibrational-relaxation time is on the order of 1~s \cite{Rellergert thesis}. Therefore, molecules
falling to excited vibrational levels of the $a^{3}\Sigma_{u}^{+}$ state are trapped in
off-resonant levels. Continuous fiber lasers at 1073~nm and 1099~nm were used to repump the
molecules from the $a(1)$ to the $c(0)$ states and from the $a(2)$ to the $c(1)$ states
respectively. Molecules in the $c$ states have a chance to decay back to the $a(0)$ state and can
be used again.

\begin{figure}
\includegraphics[scale=0.32]{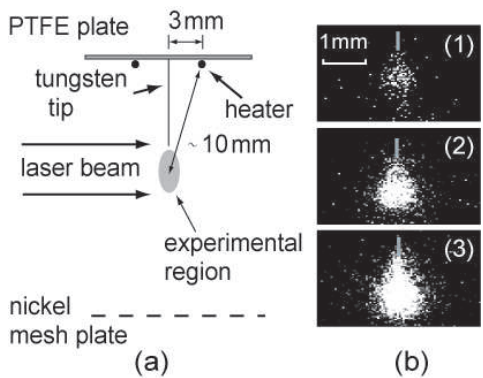}
\caption{(a)~Schematic diagram showing the setup for the molecule cloud experiment.
(b)~Fluorescence images for He$_{2}^{*}$ molecule clouds created with (1)~10~ms, (2)~30~ms and
(3)~90~ms pulse on the tip, respectively. The grey bars indicate the tip. The images are a sum of
50 camera exposures.}\label{setup1}
\end{figure}

In the first experiment, the tungsten tip was mounted at the center of a polytetrafluoroethene
(PTFE) plate (see Fig.~\ref{setup1}~(a)). The PTFE plate had a diameter of 21~mm and a thickness of
1~mm.
To make a heater, four $100~\Omega$ metal-film resistors were attached to the PTFE plate
symmetrically around the tip. A nickel mesh plate was placed 3~cm away from the PTFE plate and was
grounded. The whole device was held at the center of a helium cell with total volume of about
250~cm$^{3}$. The intensities of the fiber lasers at 1073~nm and 1099~nm were chosen to be
3~W/cm$^{2}$ and 1.5~W/cm$^{2}$ respectively. The intensity of the pulsed laser at 905~nm was
500~$\mu$J/cm$^{2}$ per pulse, and the repetition rate was 500~Hz. To create a small cloud of
He$_{2}^{*}$ molecules, a $-400$~V pulse is  delivered to the tungsten tip through a 0.1~$\mu$F
capacitor in addition to a constant voltage of $-450$~V. Electrons are emitted from the tungsten
tip when the total voltage crosses the field-emission threshold (around $-550$~V) during the pulse.
A small cloud of molecules is created near the apex of the tip as the electrons move a short
distance, lose their energy, and form bubbles in the liquid \cite{Adams}. At 2.0~K, a He$_{2}^{*}$
molecule diffuses less than 1~mm during its lifetime~\cite{McKinsey}. Thus the molecule cloud stays
together and serves as a single tracer. The size of the molecule cloud is of the order of 1~mm but
becomes larger for longer pulse durations. Typical images of a molecule cloud generated with 10~ms,
30~ms and 90~ms wide pulses are shown in Fig.~\ref{setup1}~(b). These images were taken with an
intensified CCD camera just after application of the voltage pulse to the tip. The camera was
synchronized to each laser pulse and exposed for 6~$\mu$s so as to minimize the dark current.

\begin{figure}
\includegraphics[scale=0.36]{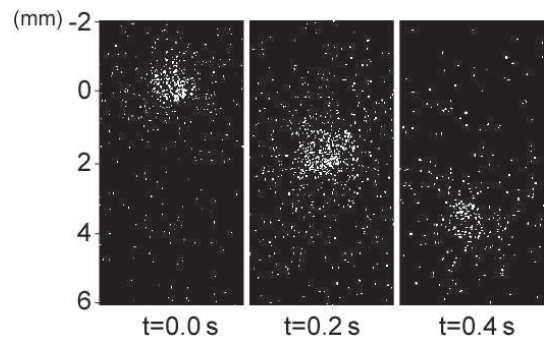}
\caption{The motion of a molecule cloud with 15V on the heater. The images were taken at 0~s, 0.2~s
and 0.4~s respectively after the cloud was created. The duration of the pulse on the tungsten tip
was 5~ms. The images are a sum of 25 camera exposures.}\label{heater on}
\end{figure}
With the heater off, the molecule cloud was observed to drift towards the nickel mesh plate. The
drift speed depended on the length of the voltage pulse on the tip. This effect results from a
transient pulling force on the normal fluid created by the moving electron bubbles \cite{Dahm}. In
order to reduce this effect but also create enough molecules for good image quality, a voltage
pulse of 5~ms duration was used. The corresponding drift velocity of the molecule cloud was about
1.8~mm/s.

As we turned on the heater, a thermal counterflow was set up in the liquid. The normal fluid flowed
away from the heater with a speed $v_n$ given in theory as \cite{Landau}
\begin{equation}
v_n=\frac{Q/A}{{\rho}ST}. \label{normal velocity}
\end{equation}
where $Q$ and $A$ are the heat power and cross-section for heat transfer; $\rho$, $S$ and $T$ are
the helium density, entropy and temperature respectively. Near the tip apex, the normal-fluid
velocity was parallel to the tip due to the geometry.
\begin{figure}
\includegraphics[scale=0.32]{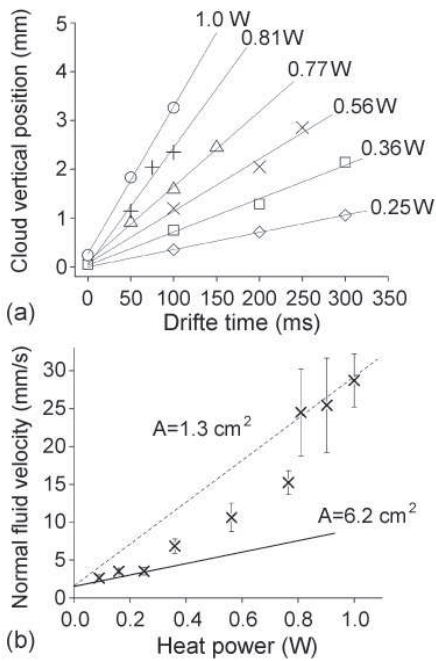}
\caption{(a)~The average vertical positions of a molecule cloud as a function of its drift time for
different heat powers. (b)~The obtained normal-fluid velocity as a function of the heat power. The
solid line and the dashed line are the theoretical curves as discussed in the text.}\label{heater
data}
\end{figure}
A typical set of images showing the motion of a molecule cloud with 15~V on the heater is shown in
Fig.~\ref{heater on}. The heater was turned on a few seconds before the molecule cloud was
generated so as to set up a steady flow of the normal fluid. The three images in Fig.~\ref{heater
on} were taken at 0~s, 0.2~s and 0.4~s respectively after the cloud was created. The number of
camera exposures for each image was chosen to be 25 in order to obtain a good signal-to-noise ratio
yet reduce image smearing. To determine the flow velocity, we fit the image of each molecule cloud
with a Gaussian function. The maximum of the Gaussian gave the center position for each cloud. For
a given drift time, several images were taken and an averaged center position was determined. In
Fig.~\ref{heater data}~(a), we show the data obtained for the averaged vertical position of each
molecule cloud as a function of its drift time. The solid lines in Fig.~\ref{heater data}~(a) are
linear fits to the data. The slopes of those solid lines give the corresponding flow velocities. In
Fig.~\ref{heater data}~(b), we plot the normal-fluid velocity obtained as a function of the heat
power. For low heat power, the normal fluid was believed to be in the laminar flow regime. Heat was
transferred to all directions below the PTFE plate. The cross-section for heat transfer in this
case was estimated to be about 6.2~cm$^2$. The solid line in Fig.~\ref{heater data}~(b) shows the
theoretical curve based on Eq.~(\ref{normal velocity}). However, as one can see, the measured data
starts to deviate from the theoretical curve when the heat power is above roughly 0.25~W. If we
take the typical length scale for the flow to be 1~cm, then the measured fluid velocity (3~mm$/$s)
gives a Reynolds number as high as 3000. It is likely that the normal-fluid flow started to become
turbulent and caused a change in heat transfer pattern. When the heat power is higher than 0.8~W,
the turbulent flow in the normal-fluid may be fully developed and the dispersion of the measured
flow velocity is large. The dashed line in Fig.~\ref{heater data}~(b) shows the theoretical curve
assuming an effective heat transfer cross-section of 1.3~cm$^2$. A smaller effective heat transfer
cross-section means most of the heat is transferred along the tip direction, for which no good
explanation has yet been found.

In the second experiment, we created a continuous molecular beam and selectively imaged a small
group of molecules which were tagged using the first excited vibrational level of the
$a^{3}\Sigma_{u}^{+}$ electronic state. To create the molecular beam, we ran the tungsten tip in a
continuous mode by applying a DC voltage higher than the field-emission threshold.\begin{figure}
\includegraphics[scale=0.26]{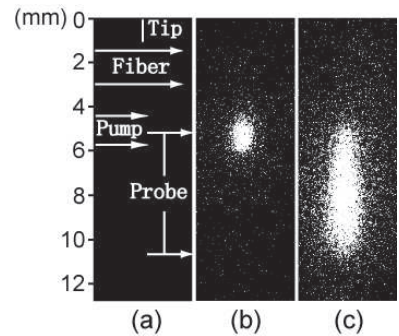}
\caption{(a)~Schematic diagram showing the lasers used in the molecule tagging experiment. (b) and
(c) show the molecule fluorescence images taken with pump laser alone and probe laser alone,
respectively. Both the pump and the probe lasers were tuned to 905~nm in order to show the beam
sizes and positions of the lasers.} \label{a1 setup}
\end{figure}
\begin{figure}
\includegraphics[scale=0.32]{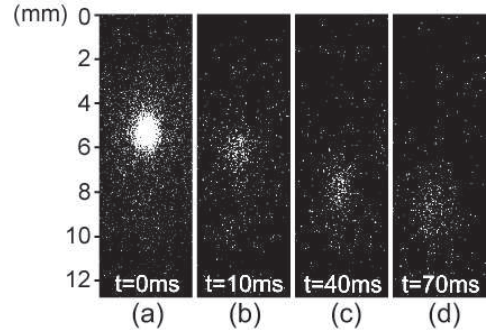}
\caption{Fluorescence images showing the positions of a small group of $a(1)$ molecules at
different delay time after they were created. The delay time between the pump and probe laser
pulses is (a)~0~ms, (b)~10~ms, (c)~40~ms and (d)~70~ms. The DC voltage on the tungsten tip was
805V.}\label{a1 motion}
\end{figure}
The field-emission current was controlled to be less than 2.5~nA to keep the electric heating
negligible. The emitted electrons moved from the tip to the nickel mesh plate leading to a
continuous pulling force on the normal fluid. A normal fluid jet was formed from the tip to the
nickel mesh plate carrying the He$_{2}^{*}$ molecules along~\cite{Dahm}. Molecules created by
field-emission initially occupy the $a(0)$, $a(1)$, and $a(2)$ excited states. To prepare a pure
population of $a(0)$-state molecules for tagging and eliminate background signal for selective
imaging, the 1073~nm and 1099~nm fiber lasers were used to illuminate a small volume near the tip
and drive molecules from the $a(1)$ and $a(2)$ excited vibrational levels into the $a(0)$ state.
Then, as shown in Fig.~\ref{a1 setup}, a focused pump laser at 910~nm was used to tag He$_{2}^{*}$
molecules by driving population from the $a(0)$ to the $c(0)$ state and relying on redistribution
of the $c(0)$ population into the long-lived $a(1)$ state (see Fig.~\ref{laser}) via non-radiative
transitions which naturally occur in a few nanoseconds \cite {Rellergert thesis}. An expanded probe
laser at 925~nm was then used to selectively image the tagged molecules by driving the $a(1)$
population into the $d$ state and inducing 640~nm fluorescence via $d\rightarrow{b}$ radiative
decay.

In Fig.~\ref{a1 motion}, we show images for a group of tagged $a(1)$ molecules taken at pump-probe
delay time of 0~ms, 10~ms, 40~ms and 70~ms respectively with 805~V on the tip. Both the pump laser
and probe laser had a pulse energy of 5~mJ and repetition rate of 10~Hz.
\begin{figure}
\includegraphics[scale=0.35]{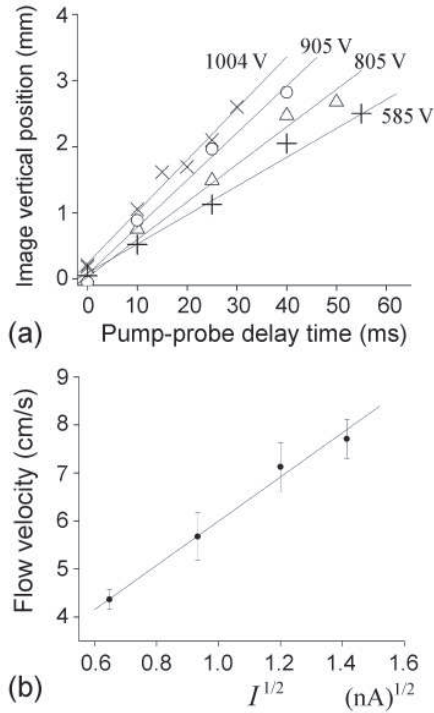}
\caption{(a)~The vertical position of the $a(1)$ molecule cloud as a function of the pump-probe
delay time at several different DC voltages on the tip. (b)~Obtained normal-fluid velocity as a
function of the square root of the current measured on the nickel mesh plate.}\label{a1 data}
\end{figure}
At each fixed pump-probe delay time, the camera was exposed ten times to obtain a single image with
a good signal-to-noise ratio. The bright image obtained with zero delay time resulted from a
two-photon transition induced by the pump laser alone at 910~nm \cite{Rellergert thesis}. We also
tested another way of producing $a(1)$ molecules by tuning the pump laser to 805~nm to drive the
$a(0)$ molecules to the $a(1)$ state through the $c(1)$ state. The signal strength obtained this
way is comparable to the one with the pump laser tuned to 910~nm. In Fig.~\ref{a1 data}~(a), we
show the vertical position of the tagged $a(1)$ molecules as a function of the pump-probe delay
time. The solid curves in Fig.~\ref{a1 data}~(a) are the linear fits to the data, and their slopes
give the corresponding flow velocity. The total driving force on the normal fluid exerted by the
moving electron bubbles is proportional to the electric current $I$ \cite{Dahm}. In steady state
the driving force on the jet is balanced by the drag force coming from the neighboring normal
fluid. If we take the typical length for the jet flow to be 1~mm (the width of the jet), the
Reynolds number is estimated to be ${\sim}5\times10^3$. The flow should be in the turbulent regime,
hence a drag force proportional to the square of the flow velocity is expected \cite{Landau}. In
Fig.~\ref{a1 data}~(b), the obtained flow velocity is plotted as a function of $I^{1/2}$. A linear
dependence is observed.

A similar result was discussed in Mehrotra's paper~\cite{Dahm}. In their experiments, a pair of
mesh grids were placed right in front of the tip to block the electric current while another pair
was placed some distance away to detect the He$_{2}^{*}$ molecules. They pulsed their tip on for
about a second and then measured the time of flight of the neutral molecules to determine the
average drift speed. Compared to their method, our technique has many advantages. For example, we
can measure the flow velocity in the steady state with the tip running all the time and map out the
velocity field along the jet.

In conclusion, we have developed practical techniques to trace the normal-fluid component in
superfluid $^{4}$He using metastable He$_{2}^{*}$ molecules. Interesting hydrodynamic phenomena in
the normal fluid were observed in the two demonstration experiments using these techniques. The
ability to track the true normal-fluid flow provides direct understanding of the hydrodynamics of
the normal-fluid component in superfluid $^{4}$He, which will in turn feed into a better
understanding of this unique two fluid system.

\end{document}